\documentclass[oneside]{article}

\usepackage{amsmath}
\usepackage{amsbsy}
\usepackage{amsfonts}
\usepackage{amssymb}
\usepackage{textcomp}
\usepackage{array}
\usepackage{graphicx}
\usepackage{bm}
\usepackage{framed}
\usepackage{bm}
\usepackage{upgreek}
\usepackage{mathrsfs}
\usepackage{hyperref}
\usepackage{color}

\newcommand{\hab}{\mathbf{h}_a}
\newcommand{\hmb}{\mathbf{h}_s}

\newcommand{\RefShape}{\mbox{$\mbox{$\Omega $}$}}
\newcommand{\CurDiv}{\mbox{$\mbox{$\text{div} $}$}}

\newcommand{\CurCurl}{\mbox{$\mbox{$\text{curl}\;$}$}}

\newcommand\Beps{\Omega^{(\varepsilon)}}
\newcommand\eps{{(\varepsilon)}}

\definecolor{reds}{RGB}{130,36,51}
\definecolor{bluem}{RGB}{94,129,181}
\definecolor{greenm}{RGB}{143,176,50}
\definecolor{orangem}{RGB}{235,98,53}
\definecolor{yellowm}{RGB}{225,156,36}
\definecolor{grays}{RGB}{180,180,180}
\definecolor{green}{RGB}{0,180,0}

\usepackage{import}
\graphicspath{{figs/}}

\newtheorem{remark}{Remark}

\DeclareMathAlphabet{\mathbf}{OT1}{cmr}{bx}{it}

\begin{document}
	

\title{\vspace{-3cm} {\bf  A  nonlinear  theory  for  fibre-reinforced  magneto-elastic rods}}

\author{
Jacopo Ciambella$^1$\!\!\!\!\! \and  Antonino Favata$^1$\!\!\!\!\! \and Giuseppe Tomassetti$^{3}$
}

%

\maketitle

\vspace{-1cm}
\begin{center}
	{\small
		
		$^2$ Department of Structural and Geotechnical Engineering\\
		Sapienza University of Rome, Rome, Italy\\
		\href{mailto:jacopo.ciambella@uniroma1.it}{jacopo.ciambella@uniroma1.it}\\[8pt]
		\href{mailto:antonino.favata@uniroma1.it}{antonino.favata@uniroma1.it}\\[8pt]

		$^3$ Roma Tre University, Engineering Department, Rome, Italy
		\href{mailto:giuseppe.tomassetti@uniroma3.it}{giuseppe.tomassetti@uniroma3.it}
	}
\end{center}

\pagestyle{myheadings}
\markboth{J.~Ciambella, A.~Favata, G.~Tomassetti}
{A  nonlinear  theory  for  fibre-reinforced  magneto-elastic rods}

\vspace{-0.5cm}
\section*{Abstract}

We derive a  model for the finite motion of a magneto-elastic rod reinforced with isotropic (spherical) or anisotropic (ellipsoidal) inclusions. The particles are assumed weakly and uniformly magnetised, rigid and firmly embedded into the elastomeric matrix.
We deduce closed form expressions of the quasi-static motion of the rod in terms of the external magnetic field and of the body forces. The dependences of the motion on the shape of the inclusions, their orientation, their anisotropic magnetic properties and the Young modulus of the matrix are analysed and discussed. Two case studies are presented in which the rod is used as an actuator suspended in a cantilever configuration. This work can foster new applications in the field of soft-actuators.

\vspace{1cm}
\noindent {\bf Keywords}: Magneto-rheological elastomers, magnetic rods, elastica, instability, soft-actuators


	\section{Introduction}
Magneto-Rheological Elastomers (MREs) are a class of functional materials whose mechanical properties can be controlled upon the application of an external magnetic field by dispersing into a non-magnetic soft matrix, magnetic hard particles. 
The use of magnetic field to achieve the actuation offers several advantages over other type of actuation such as remote and contactless control as well as the fact that it does not produce any polarization of the media nor chemical alteration \cite{Rikken2014}. 
The magnetization of the reinforcing particles by the applied field and the subsequent dipolar interactions give rise to an overall deformation that is amplified by the low elastic modulus of the matrix, usually of the order of $10$ MPa or less, and by the high susceptibility of the magnetic particles. This effect is usually referred to as \emph{huge magnetostriction} \cite{Kovetz2008}, but must not be confused with the magnetostriction of ferromagnetic crystals, as the underlying physical mechanism is completely different\footnote{In ferromagnetic materials, magnetostriction is caused by the deformation of the crystal lattice whereas in MREs the deformation can either be caused by the particle-particle interactions or by the dipolar interactions with the external magnetic field.}.

Several type of magnetic particles are nowadays commercially available including \emph{ferromagnetic}, \emph{paramagnetic} or \emph{diamagnetic} fillers \cite{Rikken2014}. Differences in their atomic nature convey different macroscopic responses: ferromagnetic and paramagnetic particles align parallel to the external magnetic field, whereas diamagnetic particles align perpendicularly \cite{Shine1987,Ciambella2017}; such a different behaviour has been exploited in a number of applications \cite{Zrinyi1996,Kimura2004,Kimura2012}. Ferromagnetic materials offer the further advantage of susceptibilities several order of magnitude higher than those of paramagnetic or diamagnetic substances and this lowers the magnetic field necessary to achieve the actuation of the samples. This is one of the reasons that makes the ferromagnetic carbonyl iron powder one of the most employed fillers in MREs \cite{Zrinyi1996,VonLockette2011,Li2011,Danas2012,Seffen2016}.

Owing to such a large availability of filler types and shapes, MREs provide a much larger design space compared to other type of soft actuators but yet require models able to account for all these features. Upon final cure, the rigid particles are locked in place into the elastomeric matrix, and the composite possess a high degree of flexibility combined with tunable stiffness that makes it capable of bearing large deformations. Moreover, if an external magnetic field is applied during the elastomer cross-linking process, the induced magnetization of the reinforcing particles makes them orient along the field lines in a chain-like structure which in turn makes the cured composite transversally isotropic. Therefore, proper models need to be formulated in the framework of large strain transversally isotropic elasticity coupled with magneto-statics.

Some of the existing theoretical studies account for the micro-geometry of the composite by evaluating the dipole interactions between adjacent particles assembled in a chain-like structure \cite{Jolly1996} or randomly dispersed \cite{Borcea2001}. However, strong kinematics assumptions, \textit{i.e.}, uniaxial deformation in the former, small strains in the latter, are made to obtain closed form expressions of the stress in terms of the magnetic quantities. 

The continuum approach to magneto-elastic response of solids dates back to the 50s with the pioneering work by Truesdell, Toupin and Tiersten (see \cite{Dorfmann2014} and references therein). These works used a direct approach to formulate the equilibrium equations based on the conservation laws of continuum mechanics. Such an approach has the advantage to making possible the coupling between magneto-elasticity with other evolutionary phenomena whose mathematical description is not of variational type \cite{RT2013,RT2017}. The same approach was applied by Dorfmann and Ogden \cite{Dorfmann2004} to formulate the equilibrium equation of magneto-elasticity at finite strains, and by DeSimone and Podio-Guidugli \cite{DeSimone1996} for ferromagnetic solids. Another continuum approach is the one used by Tiersten and Brown \cite{Brown1965} (see also \cite{Kovetz2008}), who deduced balance equations by minimizing the potential energy in terms of both magnetic and mechanical quantities; in this approach the equilibrium equations are obtained either as a global minimum or a saddle point of the functional depending upon the choice of the independent magnetic variables (see \cite{Ericksen2005} for a discussion about this point); the advantage of this method is that it allows to use variational techiques such as $\Gamma$-convergence to study physically-relevant asymptotic limits or contrained theories (see for instance \cite{desimone2}. A judicious choice of the potential energy was used in \cite{Rudykh2013,Goshkoderia2017} to study the microstructure evolution in transversally isotropic MREs (see also \cite{Galipeau2014}). Kankanala and Triantafyllidis \cite{Kankanala2004} reconciliated the two approaches, direct and variational, by showing that they yield the same governing equations and boundary conditions if the proper independent magnetic variables are chosen.

A variational formulation was more recently used in \cite{Ethiraj2016} to derive a micromechanically informed continuum model of MREs that uses a isotropic network model for polymers and extends it to the anisotropic magneto-elastic response. A reduced order model for a MR membrane was introduced in \cite{Barham2007} by exploiting the variational approach and assuming uniform and weak magnetization of the reinforcing spherical particles. These assumptions allowed the reduction of the integro-differential equations of the general theory, mechanical equilibrium and Maxwell's equations, to a set of differential equations at each material point.

In the past twenty years, a number of experiments have been carried out on MREs. Zr\'inyi and coworkers have produced and tested several type of magneto-active materials including polymer gels \cite{Zrinyi1996,Szabo1998} and elastomers \cite{Varga2006} highlighting phenomena such as magnetostriction, microscopic instabilities \cite{Varga2006} as well as macroscopic instability \cite{Szabo1998}. Von Lockette et al. \cite{VonLockette2011} produced a silicone elastomer reinforced with spherical rigid and soft magnetic particles and studied the bending behaviour of a specimen suspended between the platelets of an electromagnets. A similar configuration was exploited by Stanier et al. \cite{Stanier2016} to study the behaviour of silicone rubber reinforced with nickel coated carbon fibres; different instability mechanisms were highlighted according to the direction of the fibres. The magnetic properties of a MRE (PDMS with carbonyl iron particles) were measured in \cite{Danas2012}, where two peculiar properties were assessed. First, the magnetisation response appears to be insensitive to the level of prestrain at which the specimen was subjected to.   
Second, the magnetisation response strongly depends on the relative orientation between the particle chains and the external magnetic field.

Based upon these experimental works, we derive, in the consistent theoretical framework of 3D variational magneto-elasticity, the governing equations for the finite motion of a magneto-elastic rod reinforced with isotropic (spherical) or anisotropic (ellipsoidal) inclusions. We consider the magnetic moment in the particles as totally induced by the field, hence as susceptible to changes in the magnitude or orientation of the applied field \cite{Shine1987}. This is, indeed, different than the problem of permanently magnetized particles \cite{Singh2013,Vella2013} where the magnetization of the particle is fixed in magnitude and direction, independent of the applied field. In doing so, we consider the particles weakly and uniformly magnetised and therefore the potential energy of the system is additively decomposed into a purely mechanical term plus a part accounting for the interaction between the deformation and the applied field. The particles are further assumed rigid and firmly embedded into the elastomeric matrix, this in turns makes the demagnetization tensor dependent only on the current orientation of the particles and not on their stretch. It is further introduced an \textit{ad-hoc} choice of the susceptibility that accounts for both magnetically isotropic or anisotropic materials. These assumptions made possible to derive a closed form expression for the quasi-static motion of the rod in terms of the external magnetic field and of the body forces that act on the beam. {This approach generalises the one used in \cite{Kimura2010, VonLockette2011, Kimura2012, Stanier2016}, where only a uniform field is considered as well as incorporates the one used in \cite{Wang2015} to study the vibration of carbon-nanotubes embedded into a non-uniform magnetic field. }
It is shown that under certain conditions on the particle distribution and the applied field, the motion of the beam is governed by the classical elastica equation with forcing terms controlled by the external magnetic field. 

The structure of the paper is as follows. In Section 2, we derive the effective magneto-elastic energy of a dilute suspension of magnetic inclusions embedded into an elastic matrix. This expression is used in Section 3 to derive the energy of a rod of such a material by carrying out a formal dimensional reduction. The applications of this theory to two peculiar case studies are discussed in Section 4.

\section{The energy of an assembly of magnetic particles in a non-magnetic elastic matrix}

In this section we adopt an energetic approach to derive the effective energy  of a composite elastic body obtained by dispersing ellipsoidal magnetic inclusions into a soft, non-magnetic isotropic matrix, immersed in an applied magnetic field.

To begin with, let us fix the notation. In what follows: $\bm f:\Omega\to\Omega_c\subset{\mathscr E^3}$ is the deformation of the body from its reference configuration $\Omega$ to the current configuration $\Omega_c$, a subset of the Euclidean three-dimensional space $\mathscr E^3$; $\mathbf{X}$ is the typical point in the reference configuration whereas $\mathbf{x}$ is its image under the deformation map; accordingly $\mathbf{F}:=\partial \bm{f}/\partial \mathbf{X}=\nabla\bm f$ is the deformation gradient; $\boldsymbol a(\bm X)$ is the {\em local orientation} of the inclusions at $\bm X$ (see Fig.~\ref{fig:homogenisation}) and $\Pi$ is the volume occupied by the inclusion in the current configuration. We denote by div$(\cdot)$, grad$(\cdot)$ and curl$(\cdot)$ the divergence, gradient and curl operators with respect to the current coordinates.

The \emph{applied magnetic field} $\hab$, namely, the field that would be measured in the absence of the elastic body is the solution of the following static Maxwell's equations:
\begin{equation}
\CurDiv \hab = 0 \qquad \text{and} \qquad \CurCurl\hab = \mathbf{j}_a \qquad \text{in} \quad \mathscr E^3,
\label{Maxwellha}
\end{equation}
$\mathbf{j}_a$ being the imposed current density, which we assume to be \textit{unaffected by the presence of the body}.

Since the matrix is isotropic we assume that the strain energy of the composite  has the form $\widehat{\psi}_{\rm el}(\bm X,\bm F)=\psi_{\rm el}(\bm F,\bm a(\bm X))$ where $\psi_{\rm el}$ is an \emph{isotropic function}: $\psi_{\rm el}(\bm F\bm Q,\bm Q\bm a)=\psi_{\rm el}(\bm F,\bm a)$ for every orthogonal tensor $\bm Q$. We further make the hypothesis that the inclusions are:
\begin{enumerate}
	\item[(I)]  \emph{paramagnetic} and the intensity of the field is below the threshold that causes the saturation of the particle magnetisation;
	\item[(II)]  \emph{dilute}, so that mutual magnetic interactions can be neglected.
	\item[(III)]  \emph{firmly embedded } (\emph{cf.} \eqref{eq:12} below) in the non-magnetic soft matrix.
\end{enumerate}

With the foregoing assumptions, we shall argue in this section that the equilibrium configurations of the body are governed by the following \emph{effective energy}:
\begin{equation}\label{eq:3002}
\mathcal{E}(\bm f)=\int_\Omega \widehat{\psi}_{\rm el}\big(\bm X,\nabla\bm f\big)+\widehat{\psi}_{\rm int}(\bm f,\nabla\bm f,\bm a),
\end{equation}
where 
\begin{equation}\label{eq:2-8}
\widehat\psi_{\rm int}(\bm x,\bm F,\bm a)=-
\overline\chi\frac{\mu_0}{2}(\widehat{\bm a}_c(\bm F,\bm a)\cdot\bm h_a(\bm x))^2-\widehat{\chi}\frac{\mu_0}{2} |\hab(\bm x)|^2\quad\text{with}
\quad\widehat{\bm a}_c(\bm F,\bm a)=\frac{\bm F\bm a}{|\bm F\bm a|}
\end{equation}
is the \emph{interaction energy} between the body and the applied field. 

In \eqref{eq:2-8}, $\mu_0 = 4\pi\times 10^{-7} {\rm H}\cdot {\rm m}^{-1}$ (Henry/meter) is the \emph{magnetic permeability of vacuum}, whereas  $\overline\chi$ and $\widehat{\chi}$ are suitable \emph{effective magnetic susceptibilities} that depend on the volume fraction $\nu$ of magnetic inclusions (which we assume constant for simplicity), on the magnetic material comprising the inclusions, and on their shape (\emph{cf.} \eqref{eq:2-13} ). The vector  $\bm a_c=\widehat{\bm a}_c(\bm X,\nabla\bm f(\bm X))$ is the orientation of the inclusions in the material part that occupies the position $\bm x=\bm f(\bm X)$ in the current configuration. Thus, the effective interaction energy has both positional (since it depends on $\bm x$) and orientational (since it depends on $\bm a_c$) character.

To justify the expression \eqref{eq:2-8} for the interaction energy, we shall proceed in three steps:
\begin{itemize}
	\item [(a)] we derive the magnetic energy (see \eqref{MagnetoStaticEnergy} below) of a magnetic inclusion in terms of the applied field, of the magnetisation state of the particle and of the region $\Pi$ currently occupied by inclusion;
	\item [(b)] for $\Pi=\Pi(\bm a^c)$ a \emph{prolate spheroid}  with major axis aligned with the unit vector $\bm a^c$ immersed in a \emph{uniform applied field} $\bm h_a(\bm x)=\widehat{\bm h}_a$, we shall minimize the magnetic energy with respect to the magnetisation, and hence obtain an expression of the energy that depends only on $|\widehat{\bm h}_a|$, the intensity of the applied field, and on the relative orientation between the applied field and the \emph{current orientation} $\bm a^c$ (see \eqref{eq:9} below);
	\item [(c)] we formalize the assumption that the inclusions are firmly embedded in the matrix by prescribing the dependence of the current orientation $\bm a^c$ of the inclusion in terms of its referential orientation $\bm a(\bm X)$ and of the deformation gradient $\bm F(\bm X)$; then, we arrive at the expression \eqref{eq:2-8} for the interaction energy by a suitable volume averaging.
\end{itemize}

\subsection{A variational principle for a single particle} 
As a first step towards the construction of an averaged energy density, we focus our attention on a single inclusion immersed in an applied field $\bm h_a$. We denote by $\Pi$ the three-dimensional domain occupied by the inclusion. The magnetisation state of the inclusion is then specified by a magnetisation density $\bm m$ supported on $\Pi$, that, in turn,  generates a \emph{demagnetising field} $\bm h_s=\bm h_s\{\bm m\}$. Here we make use of curly brackets to emphasise that the dependence of the demagnetising field on $\bm m$ is non-local since $\bm h_s$ is defined as the unique square--integrable solution of the equations of magnetostatics
\begin{equation}\label{eq:22}
\begin{aligned}
&\operatorname{div}(\bm h_s+{\bm 1}_{\Pi}[{\bm m}])=0&& \text{in }\mathscr E^3,\\
&\operatorname{curl}\bm h_s=\bm 0 &&\text{in }\mathscr E^3,
\end{aligned}
\end{equation}
where ${\bm 1}_{\Pi}[\bm m]$  denotes the trivial extension of the vector field $\bm m$ to the \emph{three--dimensional space} $\mathscr E^3$:
\begin{equation}
{\bm 1}_\Pi[\bm m](\bm x)=\begin{cases}
\bm m(\bm x)&\text{if }\bm x\in\Pi,\\
\bm 0&\text{otherwise}.
\end{cases}
\end{equation}
We remark on passing out that ${\bm 1}_\Pi[\bm m]$ may have a jump at $\partial\Pi$ and hence \eqref{eq:22} should be understood in the sense of distributions. 

Under the assumption that the inclusion is paramagnetic, the magnetisation density obeys the \emph{equilibrium equation}:\footnote{This is in fact true even for a ferromagnetic material, provided that it is away from the saturation magnetisation \cite{Abbott2007}.}
\begin{equation}\label{eq:1}
\bm\Upsilon(\bm x)\bm m(\bm x)=\bm h\{\bm m\}(\bm x),\qquad \bm x\in\Pi,
\end{equation}
where $\bm\Upsilon(\bm x)$ is the \emph{inverse susceptibility} tensor at $\bm x$ (a material property) and
\begin{equation}\label{eq:7}
\bm h\{\bm m\}=\bm h_a+\bm h_{s}\{\bm m\}
\end{equation}
is the \emph{total magnetic field}.

Solving the non--local equation \eqref{eq:1} is equivalent to finding a stationary point of the \emph{magnetic--energy functional}:
\begin{equation}\label{eq:3010}
\bm m\mapsto\mathcal M(\bm m;\Pi,\bm h_a):=\mu_0\int_{\Pi}\left\{\frac 12 \bm\Upsilon\bm m\cdot\bm m-\bm h_a\cdot\bm m\right\}+\frac{\mu_0}{2}\int_{\mathscr E^3}|\bm h\{\bm m\}|^2.
\end{equation}
The density under integral sign on the right-hand side of \eqref{eq:3010} is the sum of three contributions: (i) the Helmholtz free energy density $\frac {\mu_0}2{\bm\Upsilon\bm m\cdot\bm m}$ of the particle, which accounts for the interaction between the magnetisation and the hosting lattice; (ii) the  \emph{Zeeman energy} $\mu_0\bm h_a \cdot \bm m$, which accounts for the interaction between the magnetisation and the applied field; (iii) the magnetostatic energy density $\frac{\mu_0}{2}|\bm h\{\bm m\}|^2$ of the magnetic field, whose support is the entire space, which accounts for long--range magnetic interactions \cite{Hubert1998}.

At variance with the Helmholtz and Zeeman energies, the definition of the magnetostatic energy involves an integral over the entire space. Yet,  by using  \eqref{eq:22}, it is possible to derive an alternative  expression  of the magnetic energy involving an integral extended only on the  region $\Pi$ occupied by the inclusion:
\begin{equation}\label{eq:3}
\int_{\mathscr E^3} \!\! \bm h \cdot \bm h = \int_{\mathscr E^3}\!\! \hab\cdot \hab + \int_{\mathscr E^3}\!\! \hmb\cdot \hmb +2 \int_{\mathscr E^3}\!\! \hab\cdot \hmb.
\end{equation} 
The first addendum on the right--hand side of \eqref{eq:3} is independent of the state variables and hence can be omitted from the energy calculation; moreover,  the third addendum vanishes, being the integral over the entire space of the divergence-free field $\hab$ and the irrotational field $\hmb$ \cite{Hubert1998}. By making use of \eqref{eq:22}, again it is possible to show \cite{Brown1962} that the total magnetic energy can be written as
\begin{equation}\label{MagnetoStaticEnergy}
\mathcal M(\bm m;\Pi,\bm h_a)=\mu_0\int_{\Pi}\left(\frac 12 \bm\Upsilon\bm m\cdot\bm m-\Big(\frac 12 \bm h_s\{\bm m\}+\bm h_a\Big)\cdot\bm m\right).
\end{equation}
Now, assume that the inverse susceptibility tensor is \emph{uniformly positive definite}, that is, there exists positive constant $\upsilon$ such that ${\bm \Upsilon(\bm x)\bm w\cdot\bm w}>\upsilon{|\bm w|^2}$ for every vector $\bm w$ and for every point $\bm x\in\Pi$. Then  the magnetic--energy functional \eqref{MagnetoStaticEnergy} is a convex and coercive  functional over the space of square--integrable magnetisation fields with support in $\Pi$. We can then apply the machinery of the direct method of the calculus of variations to show, by exploiting the coercivity and the quadratic structure of this functional, that there exists a unique minimizer. This minimizer is then the \textit{unique solution} of the Euler--Lagrange equation \eqref{eq:1} of the magnetic--energy functional.

\subsection{The magnetic energy as function of its current orientation}

Although finding the solution of the equilibrium equation is a linear and well--posed problem, the non-locality of the operator $\bm m\mapsto\bm h_s\{\bm m\}$ makes it difficult to find a handy expression for that solution if the shape $\Pi$ and the applied field $\bm h_a$ are arbitrary.

It is possible however, to obtain a reasonable estimate of the magnetic energy of a single particle by making a few simplifications that appear to us to be consistent with Assumptions (I)--(III) at the beginning of this section. Precisely: \\
- we let the magnetic inclusion be a prolate spheroid $\Pi(\bm a^c)$ whose major we identify with a unit vector $\bm a^c$:
\begin{equation}\label{eq:6}
\Pi=\Pi(\bm a^c),\qquad |\bm a^c|=1;
\end{equation}
\noindent - consistent with the assumption that the applied field $\bm h_a$ does not vary over the mesoscopic scale, which is larger than the typical size of the inclusion, we restrict attention to the case when the applied field is uniform:
\begin{equation}\label{eq:55}
\bm h_a(\bm x)=\widehat{\bm h}_a,\qquad \text{for all  }\bm x\in\mathscr E^3;
\end{equation}
\noindent - we assume that the particle is homogeneous and that \emph{material and shape symmetries coincide}, that is to say, the inverse susceptibility tensor is constant in $\Pi(\bm a^c)$, and given by the following expression:
\begin{equation}\label{eq:1010}
\widehat{\bm\Upsilon}(\bm a^c)=\chi_\parallel^{-1}\bm a^c\otimes\bm a^c+\chi_\perp^{-1}(\bm I-\bm a^c\otimes\bm a^c),
\end{equation}
where $\chi_\parallel>0$ and $\chi_\perp>0$ are the \emph{magnetic susceptibilities} of the material.

At this stage, we find it convenient to render explicit the dependence of the magnetic--energy functional on orientation $\bm a^c$ of the inclusion, and, on taking into account \eqref{eq:6} and \eqref{eq:55} we replace \eqref{MagnetoStaticEnergy} with:
\begin{equation}\label{MagnetoStaticEnergy2}
{\mathcal M}(\bm m;\bm a^c,\widehat{\bm h}_a)=\frac{\mu_0}2\int_{\Pi(\bm a^c)}\left(\widehat{\bm\Upsilon}(\bm a^c)\bm m\cdot\bm m-\bm h_s\{\bm m\}\cdot\bm m\right)-\mu_0\,\widehat{\bm h}^a\cdot\int_{\Pi(\bm a^c)}\bm m.
\end{equation}
It is  a standard result from magnetostatic that if the magnetisation density \emph{is constant on the ellipsoid $\Pi(\bm a^c)$}:
\begin{equation}\label{eq:44}
\bm m(\bm x)=\widehat{\bm m}\quad \text{for all}\quad\bm x\in\Pi(\bm a^c),
\end{equation}
then the restriction of the demagnetizing field in the particle is constant as well, that is,
\begin{equation}\label{eq:11}
\bm h_s\{\bm m\}(\bm x)=\widehat{\bm h}_s\quad \text{for all}\quad\bm x\in\Pi(\bm a^c);
\end{equation}
in particular, the linearity of the operator ${\bm m}\mapsto \bm h_a\{{\bm m}\}$ entails that 
\begin{equation}\label{eq:10}
\widehat{\bm h}_s=-\bm N(\bm a^c)\widehat{\bm m},
\end{equation}
where
\begin{equation}
\bm N(\bm a^c)=N_\parallel\bm a^c\otimes\bm a^c+N_\perp(\bm I-\bm a^c\otimes\bm a^c)
\label{NCur}
\end{equation}
is a positive-definite \emph{demagnetizing tensor} whose eigenvectors are collinear with the major axes of $\Pi(\bm a^c)$ \cite{bertotti}. Thus, for constant magnetisation fields having the form \eqref{eq:44} the non-local equilibrium equation reduces to an algebraic equation, namely, $(\bm\Upsilon(\bm a^c)+\bm N(\bm a^c))\widehat{\bm m}=\widehat{\bm h}_a$.

In view of the foregoing, we conclude that if \eqref{eq:55} holds, then the unique solution of \eqref{eq:1} is the constant magnetisation field $\bm m(\bm x)=\widehat{\bm m}$ with $\widehat{\bm m}$ given by:
\begin{equation}\label{eq:1001}
\widehat{\bm m}={\bm M}(\bm a^c)\widehat{\bm h}_a,\qquad \text{where}\qquad {\bm M}(\bm a^c)=(\bm\Upsilon(\bm a^c)+\bm N(\bm a^c))^{-1}.
\end{equation}
The representation formula \eqref{eq:1001} enables us to write the magnetic energy of a particle in a uniform applied field $\widehat{\bm h}_a$ as function of the orientation $\bm a^c$ only. This quantity is defined as the minimum with respect to $\bm m$ of the magnetic-energy functional 
\begin{equation}\label{eq:1007}
\widetilde{\mathcal M}(\widehat{\bm h}_a,\bm a^c):=\min_{\bm m}\mathcal M(\bm m;\bm a^c,\widehat{\bm h}_a).
\end{equation}
Since the minimizer on the right-hand side of \eqref{eq:1007} is constant, we conclude that
\begin{equation}\label{eq:9}
\widetilde{\mathcal M}(\widehat{\bm h}_a,\bm a^c)=-\text{vol}(\Pi)\frac {\mu_0}2 (\bm N(\bm a^c)+\bm\Upsilon(\bm a^c))^{-1}\widehat{\bm h}_a\cdot\widehat{\bm h}_a,
\end{equation}
where $\text{vol}(\Pi)$ is the volume of the inclusion $\Pi(\bm a^c)$. By making use of \eqref{eq:1010} and \eqref{NCur}, we can write
\begin{equation}\label{EffectiveEnergyM}
\widetilde{\mathcal M}(\widehat{\bm h}_a,\bm a^c)=-\text{vol}(\Pi)\frac {\mu_0}2 \chi\, (\bm a^c\cdot\widehat{\bm h}_a)^2-\text{vol}(\Pi)\frac{\mu_0}2 \tilde\chi|\widehat{\bm h}_a|^2,
\end{equation}
where 
\begin{equation}\label{eq:3004}
\chi=(\chi_\parallel^{-1}+N_\parallel)^{-1}-(\chi_\perp^{-1}+N_\perp)^{-1}, \qquad \tilde\chi= (\chi_\perp^{-1}+N_\perp)^{-1}.
\end{equation}
This result gives us the dependence of the magnetic energy of a single particle as a function of its current orientation $\bm a^c$. Our next step is to derive an expression for the effective energy of a dilute assembly of rigid, identical magnetic particles firmly embedded in an elastic body.

\begin{remark}{\rm 
		The expressions of $\chi$ and $\tilde{\chi}$ in \eqref{EffectiveEnergyM} can account for both the magnetic anisotropy and the shape anisotropy of the particle.  However, the origin of these two effects are remarkably different: shape anisotropy is caused by the geometry of the particle whereas magnetic anisotropy can be traced back to chemical bonds \cite{Kimura2003,Abbott2007}; for example, diamagnetic susceptibilities of the C--C bond are smaller in the direction of the bond ($\chi_\parallel$) than that normal to the bond ($\chi_\perp$), \textit{i.e.}, $\chi_\parallel<\chi_\perp<0$, that is, the anisotropic diamagnetic susceptibility defined by $\chi_a = \chi_\parallel-\chi_\perp$ is negative.
		
	}\end{remark}
	\begin{remark}{\rm  
			In the presence of two or more particles, the expression \eqref{eq:10}--\eqref{eq:11}  for the demagnetizing field should be changed to take into account the demagnetizing field generated by the magnetisation distribution outside that particle, as done in \cite{Borcea2001} in the framework of linear elasticity. The same procedure cannot be directly generalised to the case of finite deformations. However, for $d$ the diameter of a particle, and for $D$ the typical inter-particle distance, the intensity of this contribution is of the order of $(d/D)^3$, which is exactly of the same order of magnitude of  the volume fraction of magnetic particles. Accordingly, we argue that if the magnetic particles are sufficiently dilute, the mutual interaction between particles can be safely  neglected. This is indeed a first order approximation in the volume fraction as shown in\cite{Borcea2001}.
		}\end{remark}
		\begin{remark}{\rm  
				Although the presence of the body alters the total magnetic field (\emph{cf.} \eqref{eq:7}), the procedure we have used to derive the interaction energy does not require the explicit calculation of the demagnetising field.
			}\end{remark}

			\subsection{The effective interaction energy} 
\begin{figure}
				\begin{center}
					\begin{scriptsize}
						\def\svgwidth{1\textwidth}
						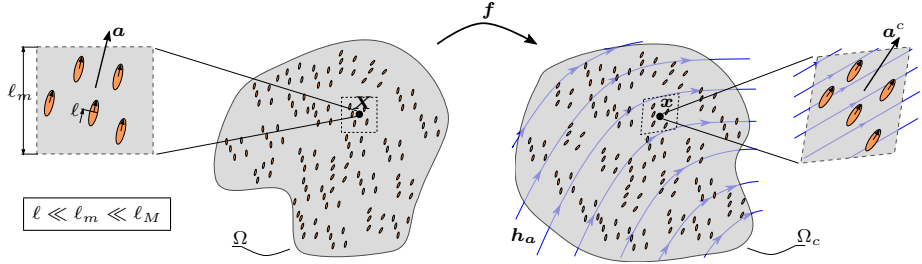
					\end{scriptsize}
				\end{center}
				\caption{Different homogenisation scales for a magneto-elastic composite reinforced with ellipsoidal inclusions: \emph{macroscale} $\ell_M$, \textit{i.e.}, $\ell_M=\int_{\RefShape}\text{d}\RefShape$, \emph{mesoscale} $\ell_m$, and \emph{characteristic length} of the inclusions $\ell$. The variation of the field $\hab$ is assumed negligible at the mesoscale $\ell_m$. The unit vector $\bm a(\bm X)$ and $\bm a^c(\bm x)$ are the common orientations of the inclusions in a mesoscopic neighbourhood of $\bm X$ and $\bm x$, respectively.\vspace{-0.7cm}}
				\label{fig:homogenisation}
			\end{figure}

			To justify our spatial averaging procedure, we make the hypothesis  that it is possible to identify a \emph{mesoscale} $\ell$ over which statistical quantities, such as volume fraction, are well defined \cite{Milton2004}. We assume that over this scale all particles appear as having constant orientation $\bm a$ and the variation of the magnetic field can be neglected at this scale (see Fig.~\ref{fig:homogenisation}). 
			This assumption allows us to define the \emph{local orientation} $\bm a(\bm X)$ and the \emph{local volume fraction} ${\nu}$ as fields in the reference configuration. 
			
			In view of our assumption (i), the current orientation $\bm a^c$ of an inclusion belonging to a mescopic neighbourhood of $\bm X$ is (\emph{cf.} $\eqref{eq:2-8}_2$):
			\begin{equation}\label{eq:12}
			\bm a^c=\frac{\bm F(\bm X)\bm a(\bm X)}{|\bm F(\bm X)\bm a(\bm X)|}.
			\end{equation}
			We now argue that the interaction energy per unit referential volume at a typical point $\bm X$ in the reference configuration is
			\begin{equation}
			{\psi}_{\text{int}}(\bm X)=\frac{{\nu}}{\text{vol}(\Pi)}\widetilde{M}\big(\bm h_a(\bm f(\bm X)),\widehat{\bm a}^c(\bm X,\bm F(\bm X))\big),
			\label{MagneticEnergy}
			\end{equation}
			namely, the product between the \emph{referential particle density} $\nu/\text{vol}(\Pi)$ and the magnetostatic energy of a single particle, with the latter given by \eqref{eq:9} with $\widehat{\bm h}_a=\bm h_a(\bm f(\bm X))$ and $\bm a^c$ given by \eqref{eq:12}. The total magnetic energy is obtained by integrating the density $\psi_{\rm int}$ over $\Omega$; on defining
			\begin{equation}\label{eq:2-13}
			\overline\chi=\nu\chi,\qquad\text{and}\qquad \widehat\chi=\nu\tilde\chi,
			\end{equation}
			and on considering the contribution of the elastic energy of the matrix $\psi_{\rm{el}}$, we arrive at \eqref{eq:3002}.

			\section{A one-dimensional model for planar rods}
			
\begin{figure}
				\begin{center}
					\begin{footnotesize}
						\def\svgwidth{.60\textwidth}
						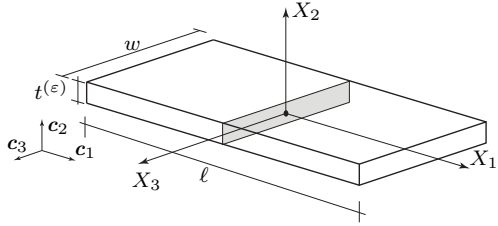
					\end{footnotesize}\vspace{-0.7cm}
				\end{center}
				\caption{Geometric properties of the thin strip studied in Sect. 3; $\ell$ is the length, $w$ the width and $t^{\eps}=\varepsilon t$ the thickness.\vspace{-0.7cm}}
				\label{fig:strips}
			\end{figure}

			In this section we consider a thin strip $\Omega^\eps$ of length $\ell$, width $w$ and thickness $t^{\eps}=\varepsilon {t}$, where $\varepsilon$ is a small dimensionless parameter. To describe the deformation of the strip, we introduce a coordinate system $(X_1,X_2,X_3)$ as shown in Fig. \ref{fig:strips}, and we let $\{\bm c_1, \bm c_2, \bm c_3\}$ be the associated orthonormal basis. We assume that the vector $\bm a$ delivering the orientation of the inclusions depends only on $X_1$ and is contained in the plane spanned by $\bm c_1$ and $\bm c_2$. We therefore write:
			\begin{equation}
			\bm a=\bm a(X_1),\qquad \bm a\cdot\bm c_3=0.
			\end{equation}
			We restrict attention to deformations on the plane spanned by $\bm c_1$ and $\bm c_2$. Consistent with the assumption of small thickness, we write the deformation as
			\begin{equation}\label{defrod}
			\bm f(X_1,X_2)=\bm r(X_1)+X_2\bm d(X_1).
			\end{equation}
			The vectors $\bm r(X_1)$ and $\bm d(X_1)$ represent, respectively, the position and the orientation of the typical \emph{cross section} $X_1\in (0,\ell)$. We rule out axial extension and shear by requiring that
			\begin{equation}
			|\bm r'|=1, \quad \bm d=\bm c_3\times\bm r'
			\end{equation}
			where a prime denotes differentiation with respect to the coordinate $X_1$.  
			
			On observing that that $\bm d'=-\kappa\bm r'$, with $\kappa:=\bm r''\cdot\bm d$ is the \emph{curvature} of the axis, it is not difficult to see that the deformation gradient is
			\begin{equation}\label{eq:2}
			\bm F(X_1,X_2)=(1-\kappa X_2)\bm r'\otimes\bm c_1+\bm d\otimes\bm c_2+\bm c_3\otimes\bm c_3.
			\end{equation}
			Since $\{\bm r'$, $\bm d$, $\bm c_3\}$ is a positively--oriented orthogonal basis, we have that 
			\begin{equation}\label{eq:3-3}
			\bm F=\bm R\bm U\qquad\text{with}\qquad\bm R=\bm r'\otimes\bm c_1+\bm d\otimes\bm c_2+\bm c_3\otimes\bm c_3\qquad\text{and}\qquad\bm U=\bm I-\kappa X_2\bm c_1\otimes\bm c_1
			\end{equation}
			is the polar decomposition of the deformation gradient, so that, thanks to the frame indifference of the elastic energy,
			\begin{equation}\label{eq:5}
			\int_{\mathcal \RefShape^\eps}\widehat\psi_{\text el}(\bm F,\bm a)=\int_{\mathcal \RefShape^\eps}\widehat\psi_{el}(\bm U,\bm a)=w\int_0^\ell \int_{-\varepsilon t/2}^{+\varepsilon t/2}\widehat\psi_{\text el}(\bm I-\kappa(X_1) X_2\bm c_1\otimes\bm c_1,\bm a(X_1)){\text d} X_2{\text d} X_1.
			\end{equation}
			Without loss of generality, we assume that $\widehat\psi_{\text el}(\bm I,\bm a)=0$ and that the reference configuration is stress--free, so that $\partial_{\bm F}\widehat\psi_{\text el}(\bm I,\bm a)=\bm 0$.
			
			Performing a Taylor expansion of the integrand with respect to $X_2$ we obtain
			\begin{equation}\label{eq:4}
			\widehat\psi_{\text el}(\bm I-\kappa X_2\bm c_1\otimes\bm c_1,\bm a)=\frac {\kappa^2}2 \partial^2_{\bm F\bm F}\widehat\psi_{\text el}(\bm I,\bm a)[\bm c_1\otimes\bm c_1]\cdot(\bm c_1\otimes\bm c_1)X_2^2+o(X_2^2).
			\end{equation}
			Since $|X_2|<\varepsilon h/2$, we have $o(X_2^2)=o(\varepsilon^2)$. Thus, on letting
			\begin{equation}
			\widetilde E(\bm a)=\partial^2_{\bm F\bm F}\widehat\psi_{\text el}(\bm I,\bm a)[\bm c_1\otimes\bm c_1]\cdot\bm c_1\otimes\bm c_1,
			\end{equation}
			and on substituting \eqref{eq:4} into  \eqref{eq:5} and on integrating with respect to $X_2$ we arrive at
			\begin{equation}\label{psiel}
			\int_{\mathcal \RefShape^\eps}\widehat\psi_{\text el}(\bm F,\bm a)=\frac{\varepsilon^3}{2}\int_0^\ell \widetilde E(\bm a)I\kappa^2{\text d}X_1+o(\varepsilon^3),\quad\text{with}\quad I=\frac {wt ^3}{12},
			\end{equation}
			which is formally identical to the bending energy of a  non-homogeneous planar rod \cite{Antman2005}.

			Next, we turn our attention to the interaction energy. We assume that the magnetic field depends on $\varepsilon$, and that it scales as
			\begin{equation}\label{eq:3-6}
			\bm h^\eps_{\text a}=\varepsilon\bm h_{\text a}.
			\end{equation}
			It is immediately seen that \eqref{eq:3-6} guarantees that bending and interaction energies scale with the same power of $\varepsilon$. Substituting \eqref{eq:3-3} and \eqref{eq:3-6} into \eqref{eq:2-8} we obtain
			\begin{align}
			\widehat\psi_{\text{int}}(\bm x,\bm F,\bm a)=&-\varepsilon^2\frac{\mu_0}{2}\left\lbrace\overline\chi\,\frac{\Big(\bm R\bm a\cdot\hab(\bm x) -\kappa X_2(\bm c_1\cdot\bm a)\big(\bm R\bm c_1\cdot\hab(\bm x)\big)\Big)^2}{|\bm I-\kappa X_2(\bm a\cdot\bm c_1)|^2}+\widehat\chi\,|\hab(\bm x)|^2 \right\rbrace\notag\\
			&=-\varepsilon^2\frac {\mu_0}{2}\left\lbrace\overline{\chi}\,(\bm{Ra}\cdot\bm h_{\text{a}}(\bm x))^2+ \widehat{\chi}\,|\hab(\bm x)|^2\right\rbrace+o(\varepsilon^2).
			\end{align}
			Moreover, by \eqref{eq:3-6} we have
			\begin{equation}
			\bm h^\eps_{\text a}(\bm x)=\varepsilon\bm h_{\text a}(\bm f(\bm X))=\varepsilon\bm h_{\text a}(\bm r(X_1))+O(X_2),
			\end{equation}
			and hence, indeed,
			\begin{equation}
			\begin{aligned}
			\int_{\Beps}\!\!\widehat\psi_{\text{int}}(\bm f,\bm F,\bm a){{\text{d}}\bm X}=-\frac{\varepsilon^3}2\int_0^\ell\,\mu_0A\left\lbrace \overline{\chi}\,\big(\bm{Ra}\cdot\hab(\bm r(X_1))\big)^2+\widehat{\chi}\,|\hab(\bm r(X_1))|^2\right\rbrace\text{d}X_1+o(\varepsilon^3),
			\end{aligned}\label{psimag}
			\end{equation}
			with $A={wt}$. 
			By  observing that $\bm R\bm a=(\bm a\cdot\bm c_1)\bm r'+(\bm a\cdot\bm c_2)\bm c_3\times\bm r'$, and scaling back the result by letting $\varepsilon=1$,  we obtain the following 1D energy:
			\begin{equation}\label{rod_energy}
			\mathcal E^{1\textsc d}(\bm r)=\int_0^\ell \widehat\psi^{1\textsc d}_{\text{el}}(|\bm r''|,\bm a)+\widehat\psi^{1\textsc d}_{\text{int}}(\bm r,\bm r',\bm a)\,\text{d}X_1,
			\end{equation}
			where
			\begin{align}
			&\widehat\psi_{\text{el}}^{1\textsc d}(\kappa,\bm a)=\frac{1}{2}{\widetilde E}(\bm a)I\kappa^2,\\
			&\widehat\psi_{\text{int}}^{1\textsc d}(\bm x,\bm r',\bm a)=-\frac {\mu_0 A}2\left(\overline\chi(\widetilde{\bm a}_c(\bm r',\bm a)\cdot\bm h_a(\bm x))^2+\widehat{\chi}\frac{\mu_0}{2} |\hab(\bm x)|^2\right),\label{eq:3-8}
			\end{align}
			are the elastic and the \emph{interaction energy} of the rod, respectively, with 
			\begin{equation}\label{currorient}
			\widetilde{\bm a}_c(\bm r',\bm a)=(\bm a\cdot\bm c_1)\bm r'+(\bm a\cdot\bm c_2)\bm c_3\times\bm r'.
			\end{equation} 
			
			It is seen that, apart from the standard elastic contribution, the interaction energy in \eqref{eq:3-8} depends on the mutual orientation of the fibres on the center-line of the rod and the applied field $\hab$; in this respect, the only part of the deformation gradient that matters is the rotation $\bm R$. It is further seen that, for uniform fields, the latter term in the energy is an additive constant that can be neglected. 
			
			When the fibers are aligned with the axis $X_1$, the density of magnetic energy is proportional to $(\bm h_a\cdot\bm r')^2$. This is in accordance with the model proposed in \cite{Cebers2003}. More recently, a model of magneto-elastic rods undergoing buckling has been proposed in \cite{Gerbal2015}. Unlike ours, these theories are direct and not deduced from the parent three-dimensional one. In order to derive the governing equations, the authors assume  that the local magnetisation depends only on the local orientation of the rod with respect to the applied field. Moreover, it is postulated that the magnetisation  orients along the rod axis (if not strictly orthogonal to the field), its longitudinal component is constant and fully determined by the maximum value achieved in the part of the rod that is mostly aligned with the applied field, \textit{e.g.}, the free tip of the cantilever rod.

			\section{Case studies}
			
\begin{figure}[h]
				\begin{center}
					\begin{tiny}
						\def\svgwidth{.5\textwidth}
						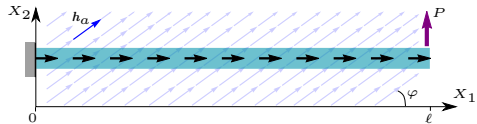
					\end{tiny}
					\caption{A soft robotic arm in its reference configuration. Dark arrows represent the fibre orientation, \textit{i.e.}, the vector field ${\bm a}$, whereas the light blue arrows are indicative of the applied field $\hab$ which forms an angle $\varphi$ with the $X_1$-axis.}
					\label{fig:robotic_arm}
				\end{center}
			\end{figure}

			As an application of the theory developed in previous sections, we derive and solve the governing equation of the cantilever  shown in Fig.~\ref{fig:robotic_arm}. The rod, subject to a dead vertical load at its free end, is immersed in a \emph{uniform} magnetic field
			\begin{equation}
			\bm h_a=H\cos\varphi\,\bm c_1+H\sin\varphi\,\bm c_2,\qquad H>0.
			\end{equation}
			This setup may be regarded as describing a prototype of a \emph{robotic arm}, which might be used to move the applied load by modulating the applied field. 
			
			We introduce the dimensionless quantities 
			\begin{equation*}
			s=X_1/\ell\in(0,1), \qquad \tilde{\bm r}(s) = \bm r(\ell\,s)/\ell,
			\end{equation*} 
			that represent the \emph{arch-length coordinate} and the parametric curve that describes the axis of the rod in its typical configuration, respectively. Since the rod is inextensible, we can adopt the following representation 
			\begin{equation}\label{eq:901}
			\tilde{\bm r}'(s)=\cos\vartheta(s)\bm c_1+\sin\vartheta(s)\bm c_2
			\end{equation}
			for the derivative of the curve at the typical point $s$, and we can express the curve in question as 
			\begin{equation}\label{eq:902}
			\tilde{\bm r}(s)=(1+u(s))\bm c_1+v(s)\bm c_2,
			\end{equation}
			with
			\begin{equation}\label{displ}
			u(s)=\int_0^s(\cos\vartheta(\bar s)-1)\text d\bar s\quad \text{and}\quad v(s)=\int_0^s\sin\vartheta(\bar s)\text d\bar s,
			\end{equation}
			respectively, the horizontal and the vertical dimensionless displacement.
			
			In the following, we assume that in the reference configuration the magnetic inclusions are parallel to the $X_1$ axis; thus, we set
			$\bm a=\bm c_1$.
			We use \eqref{rod_energy} to evaluate the contribution to the total energy coming from the interaction between the body and the applied field. In doing so, we observe that since the magnetic field is uniform, the second term on the right--hand side of \eqref{eq:3-8}, which is purely positional, can be disposed of. We also notice that, by \eqref{eq:901} we have $|\bm \tilde{r}''|^2=|\vartheta'|^2$, and that, by \eqref{currorient}, the current inclusion orientation is
			$
			\bm a^c(s)=\widetilde{\bm a}^c(\tilde{\bm r}'(s),\bm c_1)= \tilde{\bm r}'(s).
			$
			Accordingly, the 1D energy defined in \eqref{rod_energy}, when expressed in terms of the angle $\vartheta$, takes the form
			\begin{equation}\label{strip_energy}
			\widehat{\mathcal E}_{1\textsc{d}}(\vartheta)=\frac 12 \int_0^1 \left\{\frac{EI}{\ell}|\vartheta'|^2 -\mu_0 \, \overline{\chi}A\,\ell\, H^2(\cos(\vartheta-\varphi))^2\right\}\text{d}s
			\end{equation}
			where $E=\widetilde E(\bm c_1)$ is the \emph{effective Young modulus}.
			
			Now, the total energy governing equilibria of the cantilever is 
			\[
			\widehat{\mathcal E}_{\textsc{tot}}(\vartheta)=\widehat{\mathcal E}_{\textsc{1d}}(\vartheta)+\widehat{\mathcal E}_{\textsc l}(\vartheta)
			\]
			where
			\[
			\widehat{\mathcal E}_{\textsc l}(\vartheta)=-Pv(1)=-P\int_0^1\sin\vartheta(s){\rm d}s
			\]
			is the \emph{potential energy} of the applied load.
			
			On introducing the dimensionless parameters
			\begin{equation}
			h^2 = \mu_0\, \, \overline{\chi}H^2\dfrac{A\ell^2}{EI},\qquad p = \dfrac{P\ell^2}{E I},
			\end{equation}
			we can write $\widehat{\mathcal E}_{\textsc{tot}}(\vartheta)=\dfrac{E I}{2\ell} \widehat{\mathcal E}(\vartheta)$, with 
			\begin{equation}
			\widehat{\mathcal E}(\vartheta)=\frac 12 \int_0^1 \left((\vartheta^\prime(s))^2-h^2(\cos(\vartheta(s)-\varphi))^2\right) \text ds- \int_0^1
			p\sin\vartheta(s)\,\text ds. 
			\label{RobArm_energy}
			\end{equation}
			We seek configurations $s\mapsto\vartheta(s)$ that render the total energy $\widehat{\mathcal E}$ stationary. Provided that it is twice--continuously differentiable, each such configuration is a solution of the following boundary--value problem:
			\begin{equation}\
			\begin{cases}
			\vartheta^{\prime\prime}(s) - h^2 \,\sin(2\vartheta(s)+2\varphi) + p\cos(\vartheta(s))=0\quad \text{for all }s\in(0,1),
			\\\vartheta(0)=0,
			\\\vartheta^\prime(1)=0.
			\label{DimLessTheta}
			\end{cases}
			\end{equation}
			In the rest of this section, we restrict our attention to two cases particularly relevant, the second case having been considered, in a  different format, in \cite{Stanier2016} where experiments have also been conducted.
			
			\subsection*{Case 1. Field aligned with the $X_1$-axis ($\varphi=0$)}
			
			The solution of the boundary value problem \eqref{DimLessTheta} is recovered in closed form only for the two extreme cases, in the absence of the field, \textit{i.e.}, $h=0$, or in the absence of the load $p=0$; all intermediate cases must be dealt with numerically.
			However, a great deal of insight on the underlying mechanics can still be gained by studying separately two regimes, one when the applied load is low, or equivalently the stiffness of the rod is high, \textit{i.e.}, $p\ll 1$ regardless of $h$, the other one when the applied field is small compared to the load, namely $\xi=h^2/p\ll 1$. We will refer to the former case as \emph{low load regime}, to the latter as \emph{low field regime}.

			\subsection{Low load regime}\label{lowload}
			
			We firstly examine the case of a low applied load $p\ll 1$, which suggests the following first order perturbation of the solution
			\begin{equation}
			\vartheta(s) = \vartheta_0(s)+p \,\vartheta_1(s)+o(\delta),
			\end{equation}
			which, when substituted into \eqref{DimLessTheta}, leads to the boundary-value problem:
			\begin{equation}
			\begin{cases}
			\vartheta_0''+p\, \vartheta_1'' - h\, (\sin(2\vartheta_0)+2\,p\,\vartheta_1\cos(2\vartheta_0)) + p \,\cos(\vartheta_0)+o(p) = 0,\\
			\vartheta_0(0)+p\,\vartheta_1(0)+o(p)=0,\\
			\vartheta_0'(1)+p\,\vartheta_1'(1)+o(p)=0,
			\end{cases}
			\label{ODElowload}
			\end{equation}
			where for the sake of conciseness, the dependence on $s$ has been left tacit.
			By equating the coefficients at the same order, a cascade of boundary-value problems is obtained, whose first two  are
			\begin{subequations}
				\begin{align}
				\text{0-th order in $p$} \qquad &\vartheta_0'' - h^2 \sin(2\vartheta_0) = 0,\qquad \vartheta_0(0)=\vartheta_0'(1)=0\label{PDEsmalla}
				\\
				\text{1-st order in $p$} \qquad & \vartheta_1'' - 2 h^2 \cos(2\vartheta_0) \vartheta_1 = - \cos(\vartheta_0),\qquad \vartheta_1(0)=\vartheta_1'(1)=0. \label{PDEsmallb}
				\end{align}
			\end{subequations}
			We observe that \eqref{PDEsmalla} coincides with the boundary-value problem governing the equilibrium of a clamped elastica subject to a traction load at its free end,  provided that the rotation is identified with $2\vartheta_0$; accordingly, \eqref{PDEsmalla} admits only the trivial solution $\vartheta_0(s)=0$.  On taking this observation into account, we deduce from \eqref{PDEsmallb} that $\vartheta_1$ solves:
			\begin{equation}\label{th1c1}
			\vartheta_1''-2\,h^2\, \vartheta_1+1=0,
			\end{equation}
			whose  solution  can be easily determined as
			\begin{equation}\label{sol1}
			\vartheta_1(s)=\frac{1}{2\,h^2}\left(1-\frac{\cosh\big(\sqrt{2}h(s-1)  \big)}{\cosh \sqrt{2}h}  \right).
			\end{equation}
			
			Using this result, it is possible to evaluate the influence of the applied magnetic field on the \emph{effective stiffness} of the rod. We define this quantity as follows:
			\begin{equation}
			\mathfrak{s}(h) :=\Big(\left.\frac{\partial}{\partial p} v_1(p,h)\right\vert_{p=0}\Big)^{-1}\,,
			\end{equation}
			where $v_1(p,h)$ is the vertical displacement of the free end. On recalling \eqref{displ}, we can compute $v_1$ up to the first order in $p$ as
			\begin{equation}
			v_1=\int_{0}^{1}\sin(p\,\vartheta_1(s)){\rm d}s\simeq p\int_{0}^{1}\vartheta_1(s){\rm d}s=\frac{p}{h^2}\left( \frac 12 - \frac{\sqrt{2}}{4\,h}\tanh(\sqrt{2}h) \right),
			\label{tipdisp}
			\end{equation}
			that gives the following expression of the effective stiffness
			\begin{equation}
			\mathfrak{s}(h)=\frac{4\,h^3}{2\,h-\sqrt{2}\tanh(\sqrt{2}h)}.
			\label{Stiffness}
			\end{equation} 
			Equation \eqref{Stiffness} gives us a figure of merit of the rod, thought as an actuator, and  can also be used to calibrate the model with experimental data (see the discussion at the end of this paper and in particular the caption of Fig.\eqref{fig:stiffness}). 
			On passing, we note that $\mathfrak{s}$ is a monotonically increasing function whose infimum is recovered when $h\to 0$. In this limit $
			\mathfrak{s}\to3$, that is exactly the (renormalised) stiffness of a cantilever subject to a small vertical load applied at the tip.

			\subsection{Low field regime}\label{lowfield}
			
			By defining the smallness parameters $\xi=h^2/p\ll 1$, the solution of \eqref{DimLessTheta} can be expanded as a power series in $\xi$.  With a slight abuse of notation, we write $\vartheta(s) = \vartheta_0(s)+\xi\,\vartheta_1(s)+o(\xi)$.\footnote{Due to the different perturbation parameter used in \eqref{ODElowload} and \eqref{ODElowfield}, the symbols $\vartheta_0$ and $\vartheta_1$  in subsections \ref{lowload} and \ref{lowfield} denote different fields.} Correspondingly, the following boundary-value problem is 
			\begin{equation}
			\begin{cases}
			\vartheta_0''+\xi\, \vartheta_1'' - p\, \xi\, (\sin(2\vartheta_0)+2\,\xi\,\vartheta_1\cos(2\vartheta_0)) + p\,(\cos(\vartheta_0)-\xi\,\vartheta_1\,\sin(\vartheta_0))+o(\xi) = 0,\\
			\vartheta_0(0)+\xi\,\vartheta_1(0)+o(\xi)=0,\\
			\vartheta_0'(1)+\xi\,\vartheta_1'(1)+o(\xi)=0.
			\end{cases}
			\label{ODElowfield}
			\end{equation}
			By equating the coefficients at the same order, the following problems are derived
			\begin{align}
			\text{0-th order in $\xi$} \qquad &\vartheta_0'' +p \cos(\vartheta_0) = 0, \qquad \vartheta_0(0)=\vartheta_0'(1)=0, \label{ODE0thbig}\\
			\text{1-st order in $\xi$} \qquad & \vartheta_1'' - p \sin(\vartheta_0) \vartheta_1 = p \sin(\vartheta_0), \qquad \vartheta_1(0)=\vartheta_1'(1)=0\,.
			\label{ODE1stbig}
			\end{align}
			
			In solving \eqref{ODE0thbig}-\eqref{ODE1stbig}, we firstly note that the 0-th order problem~\eqref{ODE0thbig} is the same as that governing the large deflection of a cantilever with a vertical load at its free end. This problem was considered, for instance, in \cite{Wang1981,Levyakov2010, Armanini2017}. Indeed, qualitative properties of the 0-th order solutions can be derived from a phase--plane analysis, by recasting the problem \eqref{ODE0thbig} into a system of two autonomous first--order differential equations written as:
			\begin{equation}\label{ODEsystem}
			\begin{cases}
			\vartheta_0'=\kappa_0,\\
			\kappa_0'=-p\,\cos\vartheta_0,\\
			\end{cases}
			\end{equation}
			where $\kappa_0=\vartheta_0'$ is the curvature. Among the solutions of \eqref{ODEsystem}, boundary conditions select those which originate on the vertical axis (${\vartheta_0(0)=0}$) and terminate on the horizontal axis (${\vartheta_0'(1)=0}$). 
			The solution is unique for $p<p_\text{crit}^{(1)}\simeq 10.33$; for $p\geq p_\text{crit}^{(1)}$ multiple solutions can be found. In particular, the critical points of the phase plane portrait  are located on the horizontal axis and can be either \textit{centres} or \textit{saddle points}: centres comprise the set $\{(\beta_k,0):\beta_k=-\pi/2+k\pi\}$, whereas saddle points the set $\{(\beta_k,0):\beta_k=\pi/2+k\pi\}$. 
			It is also easy to check that the quantity $f(\kappa_0,\vartheta_0)=\kappa_0^2/2+p\sin\vartheta_0$ is constant along each integral curves. Accordingly, along every such curve, we have: 
			\begin{equation}\label{eq:301}
			\frac{\kappa_0^2(s)}{2}+p\sin\vartheta_0(s)=p\sin\beta=\frac{\gamma^2}2,\qquad \text{where}\quad\beta=\vartheta_0(1)\quad\text{and}\quad \gamma=\kappa_0(0).
			\end{equation}
			
\begin{figure}[h]
				\begin{center}
					\begin{tiny}
						\def\svgwidth{1\textwidth}
						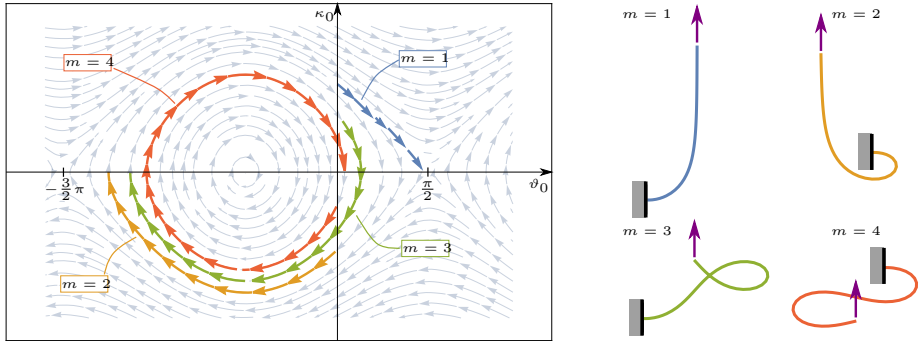
					\end{tiny}
					\caption{Phase diagram of \eqref{ODE0thbig} for $p=55\,(>p_\text{crit}^{(2)})$ with four representative solutions and the corresponding mode shapes highlighted.}
					\label{fig:PhaseDiagram}
				\end{center}
			\end{figure}

			Figure~\ref{fig:PhaseDiagram} shows the phase diagram and four representative solutions of \eqref{ODE0thbig}, and the corresponding shapes, for $p$  higher than the second critical load $p_\text{crit}^{(2)}\simeq 50.97$.\footnote{For $p>p_\text{crit}^{(2)}$ Eq.~\eqref{ODE0thbig} has indeed five solutions but the fifth is not shown in Fig.~\ref{fig:PhaseDiagram} because either its phase portrait and the shape are similar to the case $m=4$.}

\begin{figure}[h]
				\begin{center}
					\begin{tiny}
						\def\svgwidth{.7\textwidth}
						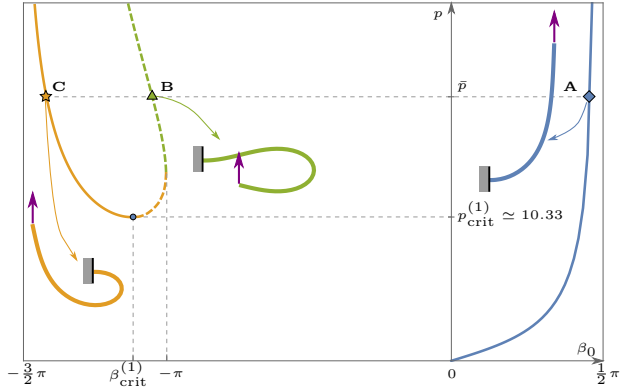
					\end{tiny}
					\caption{Bifurcation diagram for the 0-th order equation~\eqref{ODE0thbig} for $p$ against $\beta_0=\vartheta_0(1)$. The first bifurcation occurs when $p=p_\text{crit}^{(1)}\simeq 10.33$; further bifurcations are possible at higher loadings but are not shown in the graph. The dashed branch represents unstable equilibria.}
					\label{fig:p_vs_beta_0}
				\end{center}
			\end{figure}

			The above-mentioned multiplicity of solutions is further illustrated in Fig.~\ref{fig:p_vs_beta_0} where the load $p$ is plotted against the angle $\beta_0=\vartheta_0(1)$: when $p<p_\text{crit}^{(1)}=10.33$ only one  equilibrium solution of \eqref{ODE0thbig} exists and is represented by the blue branch, \textit{i.e.}, the first deformation mode, point $A$  (mode m=1 in Fig. \ref{fig:PhaseDiagram}); when ${p>p_\text{crit}^{(1)}}$, at least other two solutions are found corresponding to points $B$ and $C$ ($m=2$ and $m=3$ in Fig.~\ref{fig:PhaseDiagram}). We note that the transition between the orange and green branches occurs at $\beta_0=-\pi$ and $x(1) = 0$, characterised by null bending moment at the clamp.
			The set of solutions for $-3\pi/2<\beta_0<-\pi$, represented by dashed branch in Fig.\ref{fig:p_vs_beta_0}, are unstable equilibria \cite{Levyakov2010} and, consequently, cannot be obtained experimentally. On the other hand, the continuous branches are all stable equilibria, although those corresponding to negative $\beta_0$, \textit{i.e.}, the orange branch, are at a higher energy content and could be more difficult to attain. 
			
			Once the solution of the zero-th order equation $\vartheta_0$ is obtained, the 1-st order equation \eqref{ODE1stbig} can be solved numerically. When $p<p_{\text{crit}}^{(1)}$, the solution Eq.~\eqref{ODE0thbig} is unique and so is the solution of \eqref{ODE1stbig}. In fact, the weak version of the homogeneous equation associated to \eqref{ODE1stbig} is
			\begin{equation}
			A[\vartheta_1(s),\varphi(s)]:=\int_0^1(\vartheta'_1(s)\varphi'(s)+\gamma(s)\vartheta_1(s)\varphi(s)) {\rm d}s=0, \qquad \gamma(s):=p\sin\vartheta_0(s),
			\end{equation}
			where $\varphi(s)\in C_0^\infty([0,1])$. 
			The bilinear form $A[\vartheta_1(s),\varphi(s)]$ is continuous and, moreover,
			\begin{equation}
			A[\vartheta_1(s),\vartheta_1(s)]=\int_0^1\big((\vartheta'_1(s))^2+\gamma\vartheta_1(s)^2)\big){\rm d}s\geq C\|\vartheta_1(s)\|_{H^1([0,1])},
			\end{equation}
			which implies $A[\vartheta_1(s),\varphi(s)]$ to be coercive; thence, by the Lax-Milgram theorem, the solution of  \eqref{ODE1stbig} is unique. 
			
			\subsection*{Case 2. Field aligned with the $X_2$-axis ($\varphi=\pi/2$)}
			
			When the field is perpendicular to the fibres and no force is applied ($\varphi=\pi/2$ and $p=0$), the boundary-value problem \eqref{DimLessTheta} reduces to
			\begin{equation}
			\vartheta''+h^2 \sin(2\vartheta)=0, \qquad \vartheta(0)=\vartheta'(1)=0.
			\label{elastica}
			\end{equation}
			On observing that \eqref{elastica} can be recast into the equation governing the equilibrium of a clamped elastica subject to a compressive load, it is immediately seen that uniqueness of the solution cannot be expected \cite{cevoli}. In this sense, the external magnetic field has a \emph{destabilising effect}. 
			
\begin{figure}[h]
				\begin{center}
					\begin{tiny}
						\def\svgwidth{1\textwidth}
						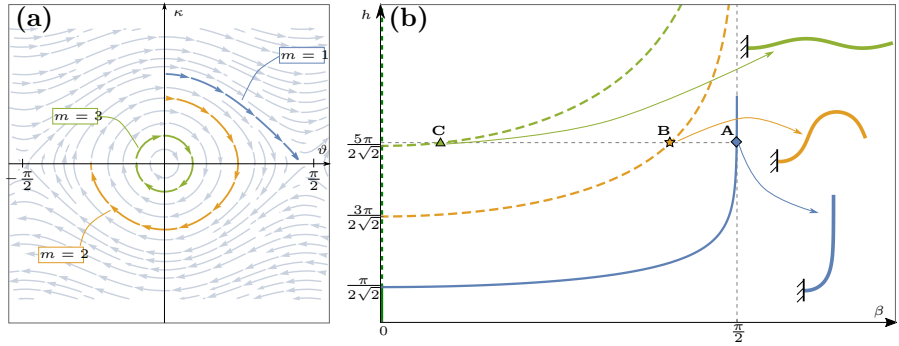
					\end{tiny}	
					\caption{(a)  Phase diagram of \eqref{phase_portrait2} in the plane $\kappa=\vartheta'$ and $\vartheta$, for $h_\text{crit}^{(3)}<h<h_\text{crit}^{(4)}$ with the first three solution trajectories highlighted. (b) Bifurcation diagram of Problem \eqref{elastica} showing the first three modes and the corresponding shapes for $h_\text{crit}^{(3)}<h<h_\text{crit}^{(4)}$. All branches have a common vertical asymptote for $\beta\rightarrow\pi/2$. The dashed branches represent unstable equilibria.}
					\label{fig:bifurcation_PhaseDiagram}
				\end{center}
			\end{figure}

			Once again, the qualitative properties of the solutions are better understood by examining the phase portrait of the system
			\begin{equation}\label{phase_portrait2}
			\begin{cases}
			\vartheta'=\kappa,\\
			\kappa'=-h^2\sin(2\,\vartheta),\\
			\end{cases}
			\end{equation}
			which is shown in Fig.~\ref{fig:bifurcation_PhaseDiagram} (a); it is noted that the symmetry in the phase portrait is due to the invariance of the solutions of \eqref{phase_portrait2} with respect to the transformation $\tilde\vartheta=-\vartheta$. Again, admissible solutions are those which originate on the vertical axis ($\vartheta(0)=0$) and terminate on the horizontal axis ($\vartheta'(1)=0$). 
			
			Equation \eqref{elastica} can be integrated once to obtain 
			\begin{equation}
			|\vartheta'|^2=h^2 \left(\cos(2\,\vartheta)-\cos(2\,\beta)\right)\,,\quad\beta=\vartheta(1).
			\label{FirstIntegral}
			\end{equation}
			which can be solved by separation of variables. A further integration of the solution between $0$ and $1$ yields an implicit relation between $\beta$ and $h$; for a given $\beta$, this relation is satisfied for
			\begin{equation}
			h=\frac{2m-1}{\sqrt{2}}\;\text{ek}(\sin(\beta)),\qquad m\in \lbrace 1,2,...\rbrace,
			\label{trascendentalbeta}
			\end{equation}
			where ek$(\cdot)$ is the complete elliptic integral of the first kind, \textit{i.e.}, ek$(q):=\int_0^{\pi/2}\left[1-q^2\sin^2(\phi)\right]^{-1/2}\text{d}\phi$ \cite{Byrd1971}. The value of the index $m$ identifies a branch in the bifurcation diagram in Fig.~\ref{fig:bifurcation_PhaseDiagram} (b). The bifurcation point of the $m$-th branch has coordinates $(0,h_\text{crit}^{(m)})$, where the critical field $h_\text{crit}^{(m)}=\frac{\pi}{2\sqrt{2}}(2m-1)$ is computed by letting $\beta\to 0$ in \eqref{trascendentalbeta}. If $h$ is below the first critical field $h_\text{crit}^{(1)}=\pi/(2\sqrt{2})$ the equation admits only the solution $\vartheta(s)=0$, \textit{i.e.}, the rod remains straight in its undeformed configuration.
			
			Each branch corresponds to a class of solution curves on the phase plane. In particular, Fig. \ref{fig:bifurcation_PhaseDiagram} (a) shows the trajectories corresponding to the three shapes A, B, and C in Fig. \ref{fig:bifurcation_PhaseDiagram} (b). It is worth noticing that the index $m$ of a branch coincides with the number of times that the solution curves associated to that branch intersect the horizontal axis.
			
			Integration of~\eqref{FirstIntegral} between 0 and $2\vartheta(s)$, $0<\vartheta(s)<\pi/2$, gives the function $\vartheta(s)$ for the first mode shape, \textit{i.e.},
			\begin{equation}
			\vartheta(s) = \arcsin\left(\sin(\beta) \;\text{sn}(s \sqrt{2} h,\sin^2(\beta))\right)
			\end{equation}
			which is expressed in terms of the Jacobi elliptic function $\operatorname{sn}(\cdot)$ \cite{Byrd1971}. By using \eqref{displ}, the horizontal $u_1$ and vertical $v_1$ displacement of the free end ($s=1$) are obtained by
			\begin{align}
			u_1 &= \frac{\pi}{2\,\text{ek}(\sin^2(\beta))}-1\\
			v_1 &= \frac{1}{\text{ek}(\sin^2(\beta))}\log\!\left(\frac{\vert \cos(\beta)\vert}{1-\sin(\beta)}\right).
			\end{align}

			\subsection*{Discussion}
			
\begin{figure}[h]
				\begin{center}
					\begin{tiny}
						\def\svgwidth{.7\textwidth}
						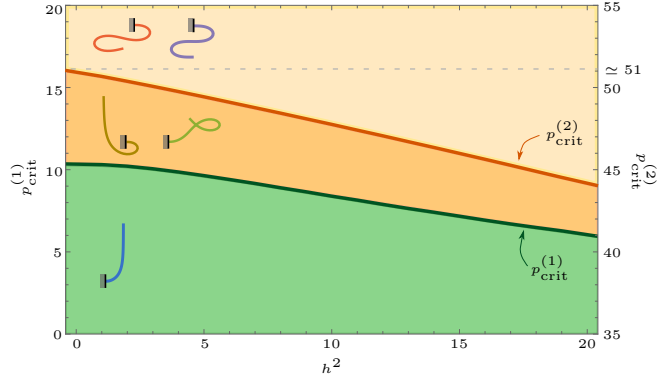
					\end{tiny}
					\caption{Phase diagram showing the multiplicity of solutions, corresponding to different colors: for $p<p_\text{crit}^{(1)}$ (green region) \eqref{DimLessTheta} has only one solution, for $p_\text{crit}^{(1)}<p<p_\text{crit}^{(1)}$ (orange region) three solutions, for $p>p_\text{crit}^{(3)}$ at least five solutions exist. The corresponding shape of the rod are drawn in the insets. Remarkably, we observe a decrease of the the critical loads $p_\text{crit}^{(1)}$ and $p_\text{crit}^{(2)}$ of \eqref{DimLessTheta} with $\varphi=0$ for an increasing magnetic field in the range $h^2\in \left[ 0,20\right]$.\vspace{-0.9cm}}
					\label{fig:pmin_vs_chi}
				\end{center}
			\end{figure}

			The possibility of using the MRE rod in the configuration shown in Fig.~\ref{fig:robotic_arm} as an actuator strongly relies on the capability of controlling its shape by modulating the applied field. As such, the appearance of multiple equilibrium configurations could be detrimental unless the transition among them can be accurately controlled or avoided.
			In this regard, the stability of the actuator with $\varphi=0$ is studied in Fig.~\ref{fig:pmin_vs_chi} by looking at the number of solutions in the $p$, $h$ plane: the green area represents the region in which \eqref{DimLessTheta} has only one solution, three solutions are found in the orange region, whereas five solutions exist in the yellow region. The continuous curves bounding the different regions are the \emph{critical loads}, that is the loads at which new solutions of \eqref{DimLessTheta} appear. It is noted that in the range $h^2\in\left[0,20 \right]$, the values of $p_\text{crit}^{(1)}$ decreases from $p_\text{crit}^{(1)}\simeq 10.33$, \textit{i.e.}, is the critical load in the absence of the field, to $p_\text{crit}^{(5)}\simeq 5$ for $h^2=20$. Such a behaviour highlights the \emph{destabilising} effect that the applied field has on the equilibrium of the rod: by increasing the field, the second deformation mode appears at lower loadings; this in turn suggests that, upon the proper control of the applied field, the transition between the second to the first mode shape can be used to realise a \emph{magnetic catapult} \cite{Armanini2017}. 
			
\begin{figure}[h]
				\begin{center}
					\begin{tiny}
						\def\svgwidth{.6\textwidth}
						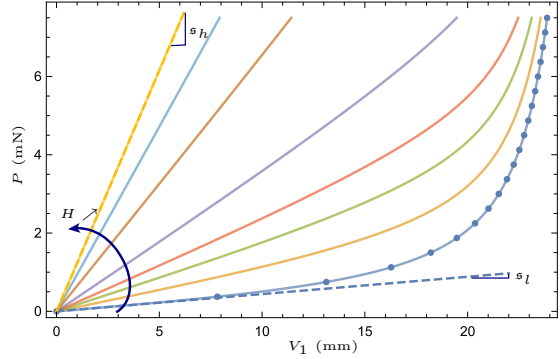
					\end{tiny}
					\caption{Load, $P$, against tip vertical displacement, $V_1$, for the configuration corresponding to mode 1 in Fig.~\ref{fig:PhaseDiagram}. The actuator properties are taken from \cite{Stanier2016}. The external magnetic field $H\in\left[0,5\right]$~kA/m produces a variation of the actuator rigidity from $\mathfrak{s}_l=4.4\times 10^{-2}$~N/m to $\mathfrak{s}_h=1.2$~N/m. The dashed lines represent the first order approximation \eqref{Stiffness}, whereas the dotted line is the solution for $h=0$ in \cite{Levyakov2010}.\vspace{-0.7cm}}
					\label{fig:stiffness}
				\end{center}
			\end{figure}

			On the other hand, if one wanted to use the actuator to lift a weight attached to its tip or move the surrounding fluid in a flap-like configuration, a quasi-static motion with the first mode shape of the configuration with $\varphi=0$ (Case 1 in previous paragraphs) would be the most effective as it would maximise, for given load and field, the displacement of the free end and at the same time would allow to continuously control the displacement of the tip by modulating the applied field. As a matter of fact, a figure of merit of an actuator is its rigidity in the operative range. For the first mode shape, when the load is low, Eq.~\eqref{Stiffness} gives the first order approximation of the rigidity. In the low field regime, the stiffness can be evaluated numerically by solving Eq.~\eqref{ODE1stbig} with $\vartheta_0$ being the first mode shape; for larger fields, the numerical solution of \eqref{DimLessTheta} can be used. The results of the calculation are plotted for $P=pEI/\ell^2$ against $V_1=\ell v_1$ in Fig.~\ref{fig:stiffness} for an actuator with length $\ell=27.5$~mm, thickness $t=3$~mm, width $w=7$~mm, $E=2.25$~MPa and $\overline{\chi}=1.32\times10^{-4}$, which are the geometric and material properties of the actuator tested in \cite{Stanier2016} made of PDMS reinforced with $6\%$ vol nickel coated carbon fibres. The dashed lines represents the first order approximation given by Eq.~\eqref{Stiffness}, the dotted line is the solution for $h=0$ reported in \cite{Levyakov2010}, whereas the continuous lines are obtained by numerically solving \eqref{DimLessTheta}. By modulating the applied field in the range $H\in\left[0, 5 \right]$~kA/m, the rigidity of the actuator can be changed by two order of magnitude from $4.4\times 10^{-2}$N/m to $1.2$ N/m; it is noted that a field value of $10~$kA/m can be easily generated by a small neodymium magnet and its below the saturation threshold of the magnetisation of the fibres\cite{Stanier2016}, thus the linear magnetic assumption  still applies.
			
\begin{figure}
				\begin{center}
					\begin{tiny}
						\def\svgwidth{1\textwidth}
						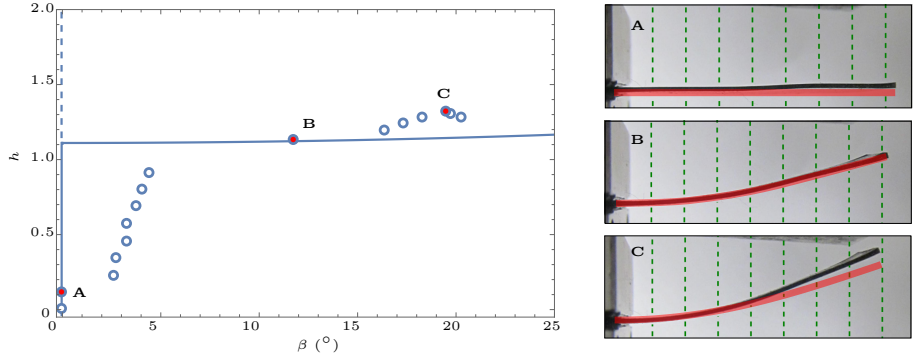
					\end{tiny}
					\caption{Applied field $h$ against the angle at the free end $\beta$ for the experimental data reported in \cite{Stanier2016} and the corresponding fitting. The insets show the shape of the actuator for the three different configurations marked as A ($h=0.12$), B ($h=1.13$) and C ($h=1.30$) in the graph. The dashed green lines indicate the direction of the homogeneous magnetic field generated in the experiment by an electromagnet.\vspace{-1cm}}
					\label{fig:fitting}
				\end{center}
			\end{figure}

			The nonlinear model of the rod with $\varphi=\pi/2$ and $p=0$ (Case 2 in previous paragraphs) is compared to the experimental data from \cite{Stanier2016} in Fig.~\ref{fig:fitting} in terms of the angle at the free end $\beta$ and the applied field $h$. The experimental data shows a sudden increase in the angle in correspondence of a critical value of the field $h_\text{crit}^{(1)}\simeq 1.11$. For such a value, the undeformed configuration of the rod $\vartheta(s)=0$, \textit{i.e.}, the trivial solution of \eqref{elastica}, becomes unstable and the system releases energy by jumping to the deformed configuration, which, in this case, has the shape of the first mode (insets A, B and C of the figure). This behaviour is due to the interplaying between the elastic bending energy and the magnetic energy in Eq.~\eqref{RobArm_energy}: by increasing the applied field, the magnetic energy of the system increases and due to the minus sign in \eqref{RobArm_energy}, the undeformed configuration passes from being a minimum of the energy to a maximum, thus the critical transition observed in the figure occurs. The nonlinear model introduced is able to describe this transition as well as the shape of the rod in the post-critical regime.
			
			\section{Conclusions and  Perspectives}
			
			The dispersion of hard magnetic inclusions into a soft matrix is a simple technique to produce soft, remotely controlled actuators that can bear large deformations.
			
			In general, the study of such structures requires the simultaneous solution of the equations governing the elastic equilibrium and the Maxwell's equations. However, we have shown that for ellipsoidal and weakly magnetised inclusions dilutely dispersed into an elastic matrix, the equilibrium of the system is governed by a reduced energy functional that depends only on the deformation and in which the magnetic field acts as a source.
			
			Starting from this result, we have derived the governing equations for the quasi-static motion of a rod-like actuator. {The model can account for large rotations/displacement of the rod, for the magnetic and shape anisotropy of the inclusions and for homogeneous and non-homogeneous external magnetic fields. As such, it is a generalisation of earlier works \cite{Kimura2012,Gerbal2015,Stanier2016}}. 
			
			Two examples have been studied with the actuator suspended in a cantilever configuration. In both cases, under the proper hypothesis, the governing equations have been partially solved in closed form and this has allowed the explicit computations of the shape of the actuator under the different regimes as well as of the critical values of the loads and the magnetic field. Different kind of instabilities were highlighted which can be hindered to exploit novel actuator configurations.
			
			The proposed nonlinear model can be extended by accounting for inertial terms that would allow the study large vibrations of slender structures embedded into a magnetic field with applications ranging from MEMS devices to carbon nanotubes.\vspace{-0.3cm}
			
			%
			\begin{footnotesize}
				
				\paragraph*{Ethics statement.} No ethical considerations apply to this work.\vspace{-0.2cm}
				
				\paragraph*{Data accessibility statement.}
				Data underlying this article are not subjected to accessibility restrictions and can be obtained upon request to the authors.\vspace{-0.2cm}
				
				\paragraph*{Competing interests statement.} We have no competing interests.\vspace{-0.2cm}
				
				\paragraph*{Authors' contributions.} All authors have contributed equally to this work.\vspace{-0.2cm}
				
				\paragraph*{Funding.} This research received no specific grant from any funding agency in the public, commercial or not-for-profit sectors.
				
			\end{footnotesize}
			


\begin{thebibliography}{9}
				
				\bibitem{Abbott2007} Abbott, J. J., Ergeneman, O., Kummer, M. P., Hirt, A. M., \& Nelson, B. J. (2007). \textit{Modeling magnetic torque and force for controlled manipulation of soft-magnetic bodies}. IEEE Transactions on Robotics, 23(6), 1247--1252. 
				
				\bibitem{Antman2005} Antman, S. S. (2005). Nonlinear problems of elasticity. Springer.
				
				\bibitem{Armanini2017} Armanini, C., Dal Corso, F., Misseroni, D., \& Bigoni, D. (2017). \textit{From the elastica compass to the elastica catapult: an essay on the mechanics of soft robot arm}. Proceedings of the Royal Society A, 473: 20160870.
				
				\bibitem{Barham2007} Barham, M., Steigmann, D. J., McElfresh, M., \& Rudd, R. E. (2007). \textit{Finite deformation of a pressurized magnetoelastic membrane in a stationary dipole field}. Acta Mechanica, 191(1--2), 1--19. 
				
				\bibitem{bertotti} Bertotti, G. (1998). \textit{Hysteresis in Magnetism: For Physicists, Materials Scientists, and Engineers}, Academic Press, Cambridge, Massachusetts.
				
				\bibitem{cevoli} Bigoni,~D. (2012). \textit{Nonlinear Solid Mechanics: Bifurcation Theory and Material Instability}. Cambridge University Press, New York, NY.
				
				
				\bibitem{Borcea2001} Borcea, L., \& Bruno, O. (2001). \textit{On the magneto-elastic properties of elastomer-ferromagnet composites}. Journal of the Mechanics and Physics of Solids, 49(12), 2877--2919. 
				
				\bibitem{Brown1962} Brown,~W.~F. (1962). \textit{Magnetostatic Principles in Ferromagnetism}. North-Holland, Amsterdam Oxford.
				
				\bibitem{Brown1965} Brown, W. F. (1965). \textit{Theory of magnetoelastic effects in ferromagnetism}. Journal of Applied Physics, 36(3), 994--1000. 
				
				\bibitem{Byrd1971} Byrd, P.F.,  Friedman, M.D. (1971). \textit{Handbook of Elliptic Integrals for Engineers and Scientists}, Second Edition, Springer--Verlag, Berlin Heidelberg New Work.
				
				\bibitem{Cebers2003} Cebers,~A. (2003). \textit{Dynamics of a chain of magnetic particles connected with elastic linkers}, J. Phys.: Condens. Matter 15, S1335-S1344. 
				
				\bibitem{Ciambella2017} Ciambella, J., Stanier, D. C., \& Rahatekar, S. S. (2017). \textit{Magnetic alignment of short carbon fibres in curing composites}. Composites Part B: Engineering, 109, 129--137. 
				
				\bibitem{Danas2012} Danas, K., Kankanala, S. V., \& Triantafyllidis, N. (2012). \textit{Experiments and modeling of iron-particle-filled magnetorheological elastomers}. Journal of the Mechanics and Physics of Solids, 60(1), 120--138. 
				
				\bibitem{desimone2} DeSimone,~A. \& James,~R. (2002). \textit{A constrained theory of magnetoelasticity}. J. Mech. Phys. Solids, 50(2):283-320.
				
				\bibitem{DeSimone1996} DeSimone, A., \& Podio-Guidugli, P. (1996). \textit{On the Continuum Theory of Deformable Ferromagnetic Solids}. Archive for Rational Mechanics and Analysis, 136, 201--233.
				
				\bibitem{Dorfmann2004} Dorfmann, A., \& Ogden, R. W. (2004). \textit{Nonlinear magnetoelastic deformations of elastomers}. Acta Mechanica, 167, 13--28.
				
				\bibitem{Dorfmann2014} Dorfmann, L., \& Ogden, R. W. (2014). \textit{Nonlinear Theory of Electroelastic and Magnetoelastic Interactions}. Boston, MA: Springer US.
				
				\bibitem{Ericksen2005} Ericksen, J. L. (2005). \textit{A Modified Theory of Magnetic Effects in Elastic Materials}. Mathematics and Mechanics of Solids, 11(1), 23--47.
				
				\bibitem{Ethiraj2016} Ethiraj, G., \& Miehe, C. (2016). \textit{Multiplicative magneto-elasticity of magnetosensitive polymers incorporating micromechanically-based network kernels}. International Journal of Engineering Science, 102, 93--119. 
				
				\bibitem{Galipeau2014} Galipeau, E., Rudykh, S., DeBotton, G., \& Ponte Casta{\~n}eda, P. (2014).\textit{ Magnetoactive elastomers with periodic and random microstructures}. International Journal of Solids and Structures, 51(18), 3012--3024.
				
				\bibitem{Gerbal2015} Gerbal, F., Wang, Y., Lyonnet, F., Bacri, J.-C., Hocquet, T., Devaud, M. (2015). \textit{A refined theory of magnetoelastic buckling matches experiments with ferromagnetic and superparamagnetic rods}. Proceedings of the National Academy of Sciences of the United States of America, 112(23):7135-7140.
				
				\bibitem{Goshkoderia2017} Goshkoderia, A., \& Rudykh, S. (2017). \textit{Stability of magnetoactive composites with periodic microstructures undergoing finite strains in the presence of a magnetic field}. Composites Part B: Engineering, 128, 19--29.
				
				\bibitem{Hubert1998} Hubert, A., \& Sch{\"a}fer, R. (1998). \textit{Magnetic Domains: The Analysis of Magnetic Microstructures}. Springer. 
				
				\bibitem{Jolly1996} Jolly, M. R., Carlson, J. D., Munoz, B. C., \& Bullions, T. a. (1996). \textit{The magnetoviscoelastic response of elastomer composites consisting of ferrous particles embedded in a polymer matrix}. Journal of Intelligent Material Systems and Structures, 7(6), 613--622. 
				
				\bibitem{Kankanala2004} Kankanala, S. V., \& Triantafyllidis, N. (2004). \textit{On finitely strained magnetorheological elastomers}. Journal of the Mechanics and Physics of Solids, 52(12), 2869--2908. 
				
				\bibitem{Kimura2003} Kimura, T. (2003). \textit{Study on the Effect of Magnetic Fields on Polymeric Materials and Its Application}. Polymer Journal, 35(11), 823--843.
				
				\bibitem{Kimura2004} Kimura, T., Yoshino, M., Yamane, T., Yamato, M., \& Tobita, M. (2004). \textit{Uniaxial alignment of the smallest diamagnetic susceptibility axis using time-dependent magnetic fields}. Langmuir, 20(14), 5669--5672. 
				
				\bibitem{Kimura2010} Kimura, T., Umehara, Y., \& Kimura, F. (2010).\textit{ Fabrication of a short carbon fiber/gel composite that responds to a magnetic field}. Carbon, 48(14), 4015--4018. 
				
				\bibitem{Kimura2012} Kimura, T., Umehara, Y., \& Kimura, F. (2012). \textit{Magnetic field responsive silicone elastomer loaded with short steel wires having orientation distribution}. Soft Matter, 8(23), 6206--6209. 
				
				\bibitem{Kovetz2008} Kovetz, A. (2008).\textit{ Electromagnetic Theory}. Oxford Univ Pr. 
				
				\bibitem{Levyakov2010} Levyakov, S. V., \& Kuznetsov, V. V. (2010). Stability analysis of planar equilibrium configurations of elastic rods subjected to end loads. Acta Mechanica, 211(1--2), 73--87.
				
				\bibitem{Li2011} Li, J., Zhang, M., Wang, L., Li, W., Sheng, P., \& Wen, W. (2011). \textit{Design and fabrication of microfluidic mixer from carbonyl iron-PDMS composite membrane}. Microfluidics and Nanofluidics, 10(4), 919--925. 
				
				\bibitem{Milton2004} Milton, G. W. (2004). The Theory of Composites. (P. G. Ciarlet, Ed.). Cambridge University Press.
				
				\bibitem{Rikken2014} Rikken, R. S. M., Nolte, R. J. M., Maan, J. C., van Hest, J. C. M., Wilson, D., \& Christianen, P. C. M. (2014). \textit{Manipulation of micro- and nanostructure motion with magnetic fields}. Soft Matter, 10(9), 1295--1308. 
				
				
				\bibitem{RT2013} Roub\'i\v{c}ek,~T. \& Tomassetti,~G. (2013). \textit{Phase transformations in electrically conductive ferromagnetic shape-memory alloys, their thermodynamics and analysis}. Arch. Rat. Mech. Analysis, 241: 1--43.
				
				\bibitem{RT2017} Roub\'i\v{c}ek,~T. \& Tomassetti,~G. (2017). \textit{Thermodynamics of magneto- and poro-elastic materials under diffusion at large strains}. ArXiV Preprint \texttt{https://arxiv.org/abs/1703.06267}.
				
				\bibitem{Rudykh2013} Rudykh, S., \& Bertoldi, K. (2013). \textit{Stability of anisotropic magnetorheological elastomers in finite deformations: A micromechanical approach}. Journal of the Mechanics and Physics of Solids, 61(4), 949--967. 
				
				
				\bibitem{Seffen2016} Seffen, K. A., \& Vidoli, S. (2016). \textit{Eversion of bistable shells under magnetic actuation: a model of nonlinear shapes}. Smart Materials and Structures, 25(6), 065010. 
				
				\bibitem{Shine1987} Shine, A. D., \& Armstrong, R. C. (1987). 
				\textit{The rotation of a suspended axisymmetric ellipsoid in a magnetic field}. Rheologica Acta
				26(2), 152--161.
				
				\bibitem{Singh2013} Singh, K., Tipton, C. R., Han, E., \& Mullin, T. (2013). \textit{Magneto--elastic buckling of an Euler beam}. Proceedings of the Royal Society A: Mathematical, Physical and Engineering Sciences, 469(May), 20130111. 
				
				
				\bibitem{Stanier2016} Stanier, D. C., Ciambella, J., \& Rahatekar, S. S. (2016). \textit{Fabrication and characterisation of short fibre reinforced elastomer composites for bending and twisting magnetic actuation}. Composites Part A: Applied Science and Manufacturing, 91, 168--176. 
				
				\bibitem{Szabo1998} Szabo, D., Szeghy, G., \& Zr{\'i}nyi, M. (1998). \textit{Shape transition of magnetic field sensitive polymer gels}. Macromolecules, 31(19), 6541--6548.
				
				\bibitem{Varga2006} Varga, Z., Filipcsei, G., \& Zr{\'i}nyi, M. (2006). \textit{Magnetic field sensitive functional elastomers with tuneable elastic modulus}. Polymer, 47(1), 227--233.
				
				\bibitem{Vella2013} Vella, D., du Pontavice, E., Hall, C. L., \& Goriely, A. (2013). \textit{The magneto--elastica: from self-buckling to self--assembly}. Proceedings of the Royal Society A, 470, 20130609. 
				
				\bibitem{VonLockette2011} Von Lockette, P., Lofland, S. E., Biggs, J., Roche, J., Mineroff, J., \& Babcock, M. (2011). \textit{Investigating new symmetry classes in magnetorheological elastomers: cantilever bending behavior}. Smart Materials and Structures, 20(10), 105022. 
				
				\bibitem{Zrinyi1996} Zr{\'i}nyi, M., Barsi, L., \& Buki, A. (1996).\textit{ Deformation of ferrogels induced by nonuniform magnetic fields}. The Journal of Chemical Physics, 104(21), 8750--8756. 
				
				\bibitem{Wang1981} Wang, C. Y. (1981). \textit{Large deflections of an inclined cantilever with an end load}. International Journal of Non-Linear Mechanics, 16(2), 155--164. 
				
				\bibitem{Wang2015} Wang, L., Liu, W. B., \& Dai, H. L. (2015). \textit{Dynamics and instability of current-carrying microbeams in a longitudinal magnetic field}. Physica E: Low-Dimensional Systems and Nanostructures, 66, 87--92. 
				
			\end{thebibliography}
\end{document}